\documentclass{aa}

\usepackage{graphicx}
\usepackage{txfonts}
\usepackage[figuresright]{rotating}
\usepackage{natbib}
\bibpunct{(}{)}{;}{a}{}{,} 
\begin{document}
   \title{The luminosity function of high-redshift QSOs - a combined analysis of GOODS and SDSS}

   \author{F. Fontanot\inst{1,2}
        S.Cristiani\inst{3}, P.Monaco\inst{1,3}, M.Nonino\inst{3},
        E.Vanzella\inst{3},
        W.N.Brandt\inst{4},
        A.Grazian\inst{5},
        J.Mao\inst{6}
        }

   \offprints{F. Fontanot}

   \institute{\inst{1}Dipartimento di Astronomia dell'Universit\`a, Via
                Tiepolo 11, I-34131 Trieste, Italy \\
                \inst{2}Max-Planck-Institut for Astronomy, Koenigstuhl 17,
                69117 Heidelberg, Germany\\                
              \email{fontanot@mpia.de} \\
                \inst{3}INAF-Osservatorio Astronomico, Via Tiepolo 11,
                I-34131 Trieste, Italy \\
                \inst{4}Department of Astronomy and Astrophysics,
                Pennsylvania State University, 525 Davey Lab,
                University Park, PA 16802 \\ 
                \inst{5}INAF-Osservatorio Astronomico di Roma, via
                Frascati 33, I-00040Monteporzio, Italy \\ 
                \inst{6}SISSA, via Beirut 2-4, I-34014 Trieste, Italy
             }

   \date{Received ...; accepted ...}

        \titlerunning{The QSO luminosity function at high-z}
        \authorrunning{Fontanot et al.}

        \abstract{\\

          {\it Aims:} In this work the luminosity function of QSOs is
          measured in the redshift range $3.5<z<5.2$ for the absolute
          magnitude interval $-21<M_{145}<-28$. Determining the
          faint-end of the luminosity function at these redshifts
          provides important constraints on models of the joint
          evolution of galaxies and AGNs.

          {\it Methods:} We have defined suitable criteria to select
          faint QSOs in the GOODS fields, checking their effectiveness
          and completeness in detail. A spectroscopic follow-up of the
          resulting QSO candidates was carried out.  The confirmed
          sample of faint QSOs is compared with a brighter one derived
          from the SDSS.  We used a Monte-Carlo technique to estimate
          the properties of the luminosity function, checking various
          parameterizations for its shape and evolution.

          {\it Results:} Models based on pure density evolution show
          better agreement with observation than do models based on
          pure luminosity evolution. However, a different break
          magnitude with respect to $z\sim 2.1$ is required at
          $3.5<z<5.2$.  Models with a steeper faint-end score a higher
          probability.  We do not find any evidence for a bright-end
          flattening at redshift $z>3.5$.

          {\it Conclusions:} The estimated space density evolution of
          QSOs indicates a suppression of the formation and/or feeding
          of supermassive black holes at these redshifts. The QSO
          contribution to the UV background is insufficient for
          ionizing the IGM at $3.5<z<5.2$.

           }

   \maketitle

\section{Introduction}
Once considered peculiar and exotic objects, quasars (QSOs) have been
recognized in recent years as an important and possibly a necessary
phase of galactic evolution, at least for spheroids (Danese et al.
2003). In particular, models of a joint evolution of galaxies and QSOs
\citep[e.g.][]{granato01,granato04,hopkins05,hopkins06} reproduce the
properties of present-day elliptical galaxies assuming that their
vigorous star formation at high redshift is quenched by the feedback
(i.e.  galactic winds). This process is triggered by the QSO activity
fed by the accretion of matter onto a supermassive black hole (SMBH)
at the center of the galaxy itself (see also Monaco \& Fontanot 2005).
In this way, the properties of high redshift QSOs and, in particular,
their luminosity function (LF) are fundamental in understanding the
phenomena driving galaxy formation. It has been hypothesized that the
feedback from AGN inverts the dark matter halo (DMH) hierarchical
sequence for the collapse of baryonic matter, a phenomenon also known
by the term {\it downsizing} \citep[e.g.][]{granato04, croton06,
  bow05, paper2, menci06, scaoh}. The depth of the potential wells in
different DMH is a key factor in determining the timescale for star
formation, accretion, and eventually for the effects of feedback,
setting the pace for AGN activity throughout cosmic history.

Great progress has recently been achieved thanks to the large amount
of data coming from major observational programs such as the {\it Two
  Degree Field QSO Redshift Survey} (2QZ, Croom et al. 2004,
www.2dFquasar.org) and the {\it Sloan Digital Sky Survey}, with the
third edition of the Quasar Catalog (DR3QSO, Schneider et al. 2005).
The 2dF-QRS contains more than 23000 QSOs typically at redshift $z \le
2.1$, while the DR3QSO lists 46420 QSOs up to $z \sim 5.4$.  However,
at high redshift the SDSS is sensitive only to the most luminous QSOs
($M_{145} \la -26.5$\footnote{Absolute AB magnitude at a wavelength of
  $145$ nanometers}). The faint-end of the high-z QSO LF, which plays
a key role in comparing different predictions of the formation and
evolution of galaxies, is left almost unconstrained due to the
challenging magnitude limit.

The recent {\it Great Observatories Origins Deep Survey} (GOODS)
provides an important database for studying QSOs with $M_{145} \la
-21$, thanks to the depth of its optical observations and to its
multi-wavelength nature. GOODS is a deep survey, not a wide one, but
it is much larger than most previous deep HST/WFPC2 surveys, covering
$320 \, arcmin^2$, 32 times the combined solid angles of the Hubble
Deep Field-North and South, and four times larger than their combined
flanking fields. The GOODS-ACS optical data were analyzed in a
previous paper (Cristiani et al. 2004 hereafter paper I) in order to
select reliable QSO candidates and make a first estimate of their
space density in the redshift interval $3.5 < z < 5.2$. The selection
was carried out by defining suitable optical selection criteria based
on magnitude limits and color criteria, and then by matching the
optical candidates with Chandra X-ray surveys (Alexander et al. 2003;
Giacconi et al. 2002).

In this paper we take advantage of the SDSS $3^{rd}$ Data Release, of
a new analysis and of a more complete spectroscopic follow up of the
GOODS data in order to determine the QSO space density at $3.5 < z <
5.2$, down to the faint-end of the LF ($M_{145} < -21$).  As in paper
I we adopt a definition of QSOs encompassing all objects with strong,
high-ionization emission lines and $M_{145} \leq -21$, including
conventional, broad-lined (type-1) QSOs, and narrow-lined, obscured
(type-2) QSOs. Throughout the paper we adopt a flat universe with
cosmological parameters $\Omega_m = 0.3$, $\Omega_{\Lambda} = 0.7$,
and Hubble constant $H_0 =70 Km \, s^{-1} Mpc^{-1}$. Magnitudes are in
the AB system.

In Sect.~2 we present the databases, the selection of the QSO
candidates, and the spectroscopic follow up; in Sect.~3 we present our
algorithm for the determination of the LF; in Sect.~4 we discuss the
results in terms of implications for current models of galaxy
formation.

\section{The database}

\subsection{Bright QSOs from the SDSS}
The bright quasar sample we use in this work was extracted from the
DR3QSO. The main SDSS catalogue covers an area of about $5282 \,
deg^2$ in photometry and $4188\ deg^2$ in spectroscopy. Photometric
catalogues were compiled with observations in $u$, $g$, $r$, $i$, $z$
bands (Fukugita et al. 1996). Several selection criteria were tailored
in order to select QSOs candidates in SDSS at different redshifts (see
e.g. Fan et al. 2001). The selection criteria have been changing
significantly over time (Schneider et al. 2005).  Richards et al.
(2002, hereafter R02) present the updated version of these selection
criteria applied to select suitable QSO candidates for spectroscopy.
We refer to these criteria in the following sections (namely to eq.~6
and~7 in R02). In particular we are interested in the selection of
high-z objects.  Due to the strategy of the SDSS observations, the
DR3QSO is a sample with an incomplete follow-up.

Fan et al. (2003) have compiled a complete sample of high-z QSOs out
of the SDSS commissioning data; however, we decided to use DR3QSO, in
order to have a larger sample of objects selected in a larger area of
the sky.  The price of this choice consists our being compelled to
make a more complex analysis to take the incompleteness of the DR3QSO
follow-up into account for our purposes (see Sect.~3).  There are 656
QSOs in DR3QSO that have a redshift between $3.5 < z < 5.2$ and that
satisfy R02 criteria. We considered those objects in the SDSS
photometric catalogue that satisfy the R02 criteria and we compared
them to the corresponding SDSS spectroscopic catalogue. This way, we
have estimated that only $\sim 21\%$ of the candidates satisfying the
R02 criteria have a spectrum.

\subsection{Faint QSOs from the GOODS}
The GOODS covers an area of $320$ arcmin$^2$, subdivided into two
$160$ arcmin$^2$ sub-fields centered on the {\it Chandra} Deep
Field-South (CDF-S) and Hubble Deep Field-North (HDF-N). The optical
data ($B_{445}$, $V_{660}$, $i_{775}$, $z_{850}$ bands) were obtained
with the Advanced Camera for Surveys (ACS) onboard {\em HST} in the
framework of the GOODS/ACS survey described in Giavalisco et al.
(2004a). The catalogues used in this paper, prepared using the
SExtractor package (Bertin \& Arnouts 1996), are based on the version
{\it v1.0} of the reduced, calibrated, stacked, and mosaiced images
acquired with HST and ACS as part of the GOODS ACS Treasury
program\footnote{see, for details
  http://www.stsci.edu/science/goods/}. The catalogues are
$z$-band-based, that is, source detection was made using the $z$-band
images. The magnitude limits in the four bands are $27.50$ ($B_{445}$
band), $27.25$ ($V_{660}$ band), $27.00$ ($i_{775}$ band), and $26.5$
($z_{850}$ band) at S/N 10 for point sources.

The HDF-N and CDF-S fields were observed in the X-rays with Chandra
for 2~Ms and 1~Ms, respectively (Alexander et al. 2003, hereafter A03;
Giacconi et al. 2002, hereafter G02), providing the deepest views of
the Universe in the 0.5--8.0~keV band.  The X-ray completeness limits
over $\approx$~90\% of the area of the GOODS fields are similar, with
flux limits (S/N$=5$) of $\approx
1.7\times10^{-16}$~erg~cm$^{-2}$~s$^{-1}$ (0.5--2.0~keV) and $\approx
1.2 \times10^{-15}$~erg~cm$^{-2}$~s$^{-1}$ (2--8~keV) in the HDF-N
field and $\approx 2.2\times10^{-16}$~erg~cm$^{-2}$~s$^{-1}$
(0.5--2.0~keV) and $\approx 1.5\times10^{-15}$~erg~cm$^{-2}$~s$^{-1}$
(2--8~keV) in the CDF-S field (A03).  The sensitivity at the aim point
is about 2 and 4 times better for the CDF-S and HDF-N, respectively.
As an example, assuming an X-ray spectral slope of $\Gamma=$~2.0, a
source detected with a flux of
$1.0\times10^{-16}$~erg~cm$^{-2}$~s$^{-1}$ would have both observed
and rest-frame luminosities of $8\times 10^{42}$~erg~s$^{-1}$ and
$3\times 10^{43}$~erg~s$^{-1}$ at $z=3$ and $z=5$, respectively
(assuming no Galactic absorption).  A03 produced point-source
catalogues for the HDF-N and CDF-S and G02 for the CDF-S.

\subsubsection{The color selection}
The selection of the QSO candidates was carried out in the magnitude
interval $22.25 < z_{850} < 25.25$. To avoid the lower quality zones
at the borders of the GOODS ACS mosaics, we carried out the selection
in a slighty reduced area of $157.1$ arcmin$^2$ in the HDF-N and
$156.4$ arcmin$^2$ in the CDF-S.

Expected QSO colors in the ACS bands were estimated as a function of
redshift using a template of Cristiani \& Vio (1990, hereafter CV90)
for the QSO spectral energy distribution (SED) convolved with a model
of the intergalactic medium (IGM) absorption (see Appendix).  The same
scheme will be adopted in the following for the estimate of expected
colors in the different SDSS passbands, unless otherwise specified.

In paper I, four optical criteria had been tailored to select QSOs at
progressively higher redshift in the interval $3.5 \la z \la 5.2$.:
{\small
\begin{eqnarray}
i-z&<&0.35 ~~{\rm AND}~~ 1.1<B-V<3.0 ~~{\rm AND}~~ V-i<1.0 ~~~~\\
i-z&<&0.35 ~~{\rm AND}~~ B-V>3.0\\
i-z&<&0.5  ~~{\rm AND}~~ B-V>2.0 ~~{\rm AND}~~ V-i>0.8 \\
i-z&<&1.0  ~~{\rm AND}~~ V-i>1.9.
\end{eqnarray}
}
As already shown in paper I, these color selection criteria are
expected to be most complete and reliable at $z > 4$ (corresponding to
criteria 2--4). These criteria select a broad range of high-$z$ AGN,
not limited to broad-lined (type-1) QSOs, and are less stringent than
those typically used to identify high-$z$ galaxies (e.g. Giavalisco et
al. 2004b). Below $z \simeq 3.5$, the typical QSO colors in the ACS
bands move close to the locus of stars and low-redshift galaxies.
Beyond $z \simeq 5.2$, the $i-z$ color starts increasing and infrared
bands would be needed to identify QSOs efficiently with an
``$i$-dropout'' technique. The full set of color criteria selects
objects in the redshift range ($3.5 < z < 5.2$), while the subset
(2-4) is able to select objects with $z>4$. To avoid contamination
from spurious sources, we limited our selection to $z_{850}$
detections with $S/N > 5$.

\subsubsection{Matching the color selection to X-ray
  catalogs}\label{xsec}
The optical candidates selected with the criteria (1-4) were matched
with X-ray sources detected by {\it Chandra} (A03, G02) within an
error radius corresponding to the $3~\sigma$ X-ray positional
uncertainty. With this tolerance, the expected number of false matches
was five and indeed two misidentifications, i.e. cases in which a
brighter optical source lies closer to the X-ray position, were
rejected (both in the CDF-S). As already shown in paper I, given the
flux limits of the Chandra surveys, Type-1 QSOs with $M_{145}<-21$
should be detectable up to $z \ga 5.2$, up to an optical-to-X-ray flux
ratio $\alpha_{ox} \ga -1.7$. We compare this result with the estimate
of $\alpha_{ox}$ statistics given by Steffen et al. (2006, their Table
5). They observe a mean value ($\alpha_{ox} = -1.408 \pm 0.165$) at
this optical luminosity, lower than our limit. Conversely, any $z>3.5$
source in the GOODS region detected in the X-rays must harbor an AGN
($L_x(0.5-2~{\rm keV}) \ga 10^{43}~{\rm erg~s^{-1}}$).

The resulting sample consists of $16$ candidates, $10$ in CDF-S and
$6$ in HDF-N (Table~\ref{new-cand}). With respect to the candidates in
paper I the following differences are found: in the HDF-N (CDF-S)
three (one) candidates selected in paper I disappear from the
selection, while three (one) new candidates enter the selection. These
differences are due to the improvement both in the photometry and in
the astrometry of the v1.0 GOODS/ACS catalogue with respect to v0.5.
In particular, a better matching is possible between the optical and
the X-ray sources. We discuss the implication of these changes in more
detail in the next section.
\begin{sidewaystable*}
\begin{minipage}[t]{\textwidth}
\caption{QSO candidates.}
\label{new-cand}
\renewcommand{\footnoterule}{}  
\centering
\begin{tabular}{cccccccccccccc}
\hline
\hline
\multicolumn{12}{c}{High-redshift QSO candidates in the CDF-S.}\\
\multicolumn{2}{c}{Optical} & \multicolumn{2}{c}{Optical -- X-ray} &
$z_{850}$ &  $B-V$ &  $i-z$ & $V-i$ &
\multicolumn{2}{c}{$F_x$\footnote{Objects with upper limits both in
the soft and hard band have been detected in other sub-bands.}} &
Flux &  spectr.&  \multicolumn{2}{c}{spectroscopic} \\   
 RA &  DEC &  $\Delta$ RA & $\Delta$ DEC &  AB &  AB &  AB & AB &
$0.5-2~{\rm keV}$ & $2-8~{\rm keV}$ &  Radius &
redshift &
\multicolumn{2}{c}{identification} \\
 $3^h+ ~m~ s$ &  $-27^{\circ}+ ~'~``$ & (arcsec) &  (arcsec) & (mag) &
(mag) & (mag) & (mag) & \multicolumn{2}{c}{$(10^{-16} {\rm
erg~s{-1}~cm^{-2}})$} & (pixels) & & &  \\ 
\hline
 32 04.93\footnote{From the supplementary X-ray catalogue by
\citet{alexander03}} &   
           44 31.7& $+1.7$ & $-0.5$ &23.668& $1.41$& $0.05$ & 0.20
&$<0.74$&$4.10$& 1.75& 3.462& QSO & \citet{sc00}\\  
 32 14.44& 44 56.6& $+0.0$ & $+0.0$ &23.094& $2.77$& $0.43$ & 1.42
&$0.41$ &$<4.3$& 3.58& 0.738 & Galaxy & \citet{szok04}\\ 
 32 18.83& 51 35.4& $+0.0$ & $+0.0$ &25.085& $2.26$& $0.16$ & 0.49
&$0.74$ &$14.0$& 2.58& 3.660& Type-2 QSO & \citet{szok04}\\  
 32 19.40\footnote{From the supplementary X-ray catalogue by
\citet{alexander03}} &   
           47 28.3& $+0.8$ & $-0.7$ &24.676& $1.29$& $0.05$ & 0.42
&$<0.31$&$1.17$& 1.96 & 3.700 & Galaxy & \citet{eros06}\\ 
 32 29.29\footnote{Detected in the X-ray by \citet{Giacconi02} and not
by \citet{alexander03}}&   
           56 19.4& $-1.0$ & $+0.1$ &24.984& $>3$  & $0.14$ & 1.72 &
$0.50$&$<7.6$& 1.31& 4.759& QSO & \citet{eros05}\\  
 32 29.84& 51 05.8& $+0.0$ & $+0.0$ &24.642& $2.36$& $-0.04$& 0.55 &
$3.06$&$31.8$& 1.97& 3.700& Type-2 QSO & \citet{Norman02}\\  
 32 39.66& 48 50.6& $+0.0$ & $+0.0$ &24.547& $3.20$& $0.21$ & 0.95 &
$7.48$&$70.6$& 2.46& 3.064& Type-2 QSO & \citet{szok04}\\  
 32 40.83& 55 46.7& $+0.0$ & $+0.0$ &25.183& $1.77$& $0.09$ & 0.79 &
$5.42$&$93.4$& 2.72 & 0.625 & Type-2 AGN & \citet{szok04}\\  
 32 41.85& 52 02.5& $+0.0$ & $+0.0$ & 22.430 & $1.46$& $-0.02$& 0.48 &
$16.7$&$38.2$& 1.25& 3.592& QSO & \citet{szok04}\\  
 32 42.83& 47 02.4& $+0.0$ & $+0.0$ &25.184& $1.43$& $0.00$ &-0.13 &
$6.31$&$16.4$& 1.22& 3.193& QSO & \citet{szok04}\\  
\hline
\hline
\multicolumn{12}{c}{High-redshift QSO candidates in the HDF-N.}\\
 $12^h+ ~m~ s$ & $62^{\circ}+ ~'~``$ \\ 
\hline
\\
 36 29.44& 15 13.2& $+0.0$ & $+0.1$ &23.701& $1.29$& $0.19$ & 0.73&
$8.05$&$11.2$& 1.21& 3.652 & QSO &\citet{cow04}\\
 36 42.22& 17 11.6& $+0.2$ & $+0.2$ &24.022& $1.28$& $0.23$ & 0.39&
$9.48$&$23.2$& 1.50& 2.724& QSO &\citet{Barger01}\\      
 36 43.09& 11 08.8& $+0.2$ & $-0.6$ &22.918& $1.11$& $0.06$ & 0.48&
$0.75$&$<2.0$& 7.13& 3.234\footnote{Multiple object: \citet{Cohen00} 
 report $z=0.299$.} & --- &\citet{cow04}\\
 36 47.96& 09 41.6& $-0.3$ & $-0.1$ &23.761& $4.81$& $0.15$ & 2.07&
$2.72$&$4.89$& 1.27& 5.186& QSO &\citet{Barger02} \\ 
 37 03.98& 11 57.8& $-0.7$ & $-0.8$ &25.067& $1.36$& $0.05$ & 0.18&
$0.37$&$3.49$& 1.38& 3.406& QSO &\citet{Barger03}\\ 
 37 23.71& 21 13.3& $+0.2$ & $+0.7$ &23.621& $1.21$& $0.02$ & 0.32&
$5.07$&$11.6$& 1.37& 3.524 & QSO &\citet{cow04} \\
\hline
\end{tabular}
\vfill
\end{minipage}
\end{sidewaystable*}

\subsubsection{Spectroscopic follow-up}\label{spectro}
Spectroscopic information for all 16 candidates has been obtained by
various sources, as listed in Table~\ref{new-cand}, particularly in
the framework of the GOODS Team Spectroscopic Survey (Vanzella et al.
2005, 2006).

Thirteen ($5$ in the HDF-N and $8$ in CDF-S) candidates turn out to be
$AGN$, with $11$ QSOs at $z > 3$ ($4$ and $7$), of which $2$ ($0$
and $2$) are identified as Type II. One object ({\sc GDS
  J033229.29-275619.5}) in particular turns out to be a QSO with a
redshift $z=4.759$. This candidate was observed in the framework of
the GOODS/FORS2 spectroscopic follow-up (Vanzella et al. 2006) and its
reduced spectrum (one-hour exposure with $FORS2$ at $ESO-VLT$) is
shown in Fig.~\ref{spfig}(a). In addition to the already known QSO
{\sc GDS J123647.97+620941.7} at $z=5.189$ (Barger et al. 2002), {\sc
  GDS J033229.29-275619.5} brings to two the total number of objects
in the redshift range $4<z<5.2$, where our selection criteria are
expected to be most complete and reliable.

If we consider the changes in the selected candidates between paper I
and the present work, we can also draw some interesting conclusions.
As shown in the previous section, $3$ objects selected in paper I are
not recovered in the present work. Two of these objects turn out to be
lower-redshift galaxies. The third object is a confirmed QSO candidate
with a redshift $z=2.573$. Our analysis shows that its new colors do
not satisfy our criteria anymore. On the other hand, if we consider
objects present both in the first and in the present selections, those
are confirmed QSOs. Finally, $3$ new objects are selected with the
updated version of the ACS-catalogue and all are confirmed QSOs. Among
them we found an additional candidate ({\sc GDS J033240.82-275041.4}),
which is a low-redshift type--2 AGN.

\subsubsection{Checking the completeness of the X-ray criterion}
\begin{figure*}[htp]
\centering
\includegraphics[width=12cm]{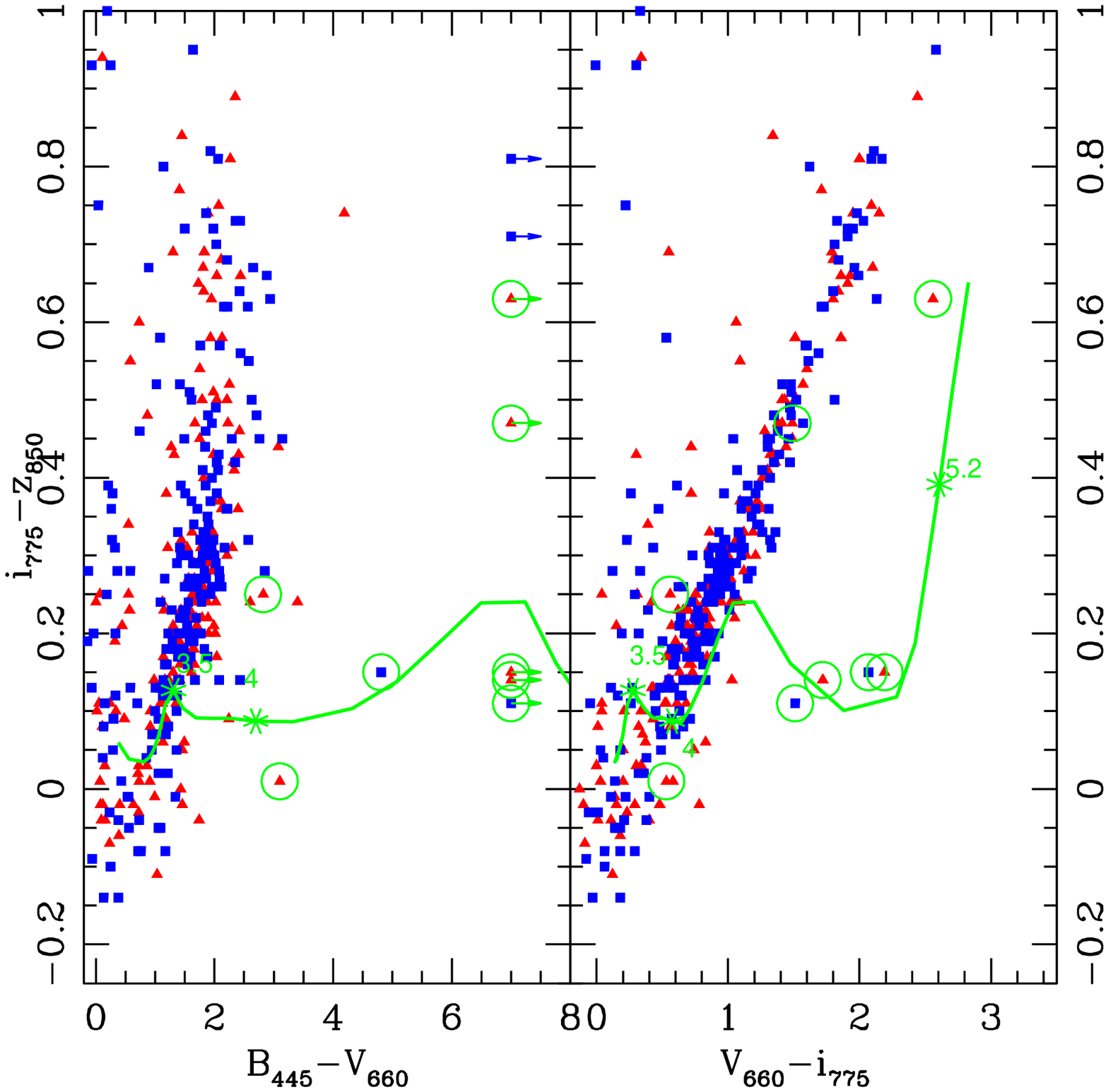}
\caption{Left Panel: the $B_{445}-V_{660}$ versus $i_{775}-z_{850}$
  color in HDF-N and CDF-S. Right Panel: the $V_{660}-i_{775}$ versus
  $i_{775}-z_{850}$ color in HDF-N and CDF-S. The color of point-like
  candidates are indicated with filled (blue) squares (HDFN) and
  filled (red) triangles (CDFS). The empty circles
  mark the positions of our $8$ final candidates in the combined
  fields. The solid line shows the locus of expected QSO colors at $3
  < z < 5.5$; we have marked with asterisks the positions
  corresponding to redshift $3.5$, $4$ and $5.2$.
  \label{vi-iz}}
\end{figure*}
It is worth noting that the $z=4.76$ QSO ({\sc GDS
  J033229.29-275619.5}) is a faint X-ray source that was only selected
by matching our optical candidates with the supplementary X-ray
catalogue by G02. This brings up the possibility that our selection
process might be missing QSOs, in particular at $z \ge 4$, if their
X-ray flux is reduced below detectable levels (e.g. by peculiar
absorption).  To assess this potential problem we replaced the X-ray
criterion (Sect.~\ref{xsec}) with a morphological one and applied it
to a subset of the color-selected candidates at $z \ge 4$.

The SExtractor software provides a set of parameters quantifying the
radius (in pixels) inside which a given percentage of the source flux
is concentrated. In the following we fix this percentage at $10\%$,
and we adopt as a diagnostic quantity the related parameter (we refer
to it as {\sc FLUX\_RADIUS}). SExtractor also provides a parameter
termed {\sc CLASS\_STAR}, that quantifies the stellar appearance of
the source. For magnitudes in the range of interest, point-like
sources ({\sc CLASS\_STAR} $\simeq 1$) and extended sources ({\sc
  CLASS\_STAR} $\simeq 0$) are segregated well, while they tend to mix
at fainter magnitudes.

The analysis of the distribution of the {\sc FLUX\_RADIUS} values in
our sample of QSO candidates shows that objects with greater {\sc
  FLUX\_RADIUS} values turn out to be either lower redshift
contaminant galaxies or Type--2 QSOs. This parameter is typically
lower than $2$ for Type--1 QSOs, while Type--2 QSOs show {\sc
  FLUX\_RADIUS} $ < 2.5$. Values greater than $2.5$ usually correspond
to low-redshift galaxies and misidentifications. It is interesting to
note that, out of the three candidates that were selected in paper I
and not in the present work (see Sect.~\ref{spectro}) the two having
{\sc FLUX\_RADIUS} $ > 2.5$ turn out to be relatively low-redshift
galaxies.  This criterion can be strengthened if we also consider the
{\sc CLASS\_STAR} parameter.  In this way a combined morphological
selection has been adopted:
{\small
\begin{eqnarray}
FLUX\_RADIUS&<&2.0 ~~{\rm AND}~~ CLASS\_STAR>0.7
\end{eqnarray}
}
To distinguish reliable QSO candidates from stars we compute the
distance $D_{col}(z)$ between the point-like candidates and the
theoretical locus of quasar colors as a function of redshift at $z \ge
4$, in the $\, (B-V\, , \, V-i \, , \, i-z) \,$ space

\begin{displaymath}
D_{col}(z)= \frac{\sqrt{\sum_{i=1}^{3}
[col^{th}_i(z)-col^{obs}_i]^2}}{\sqrt{3}}
\end{displaymath}

For $B$-band dropouts we consider the distance in the $\, (V-i \, , \,
i-z) \,$ space:

\begin{displaymath}
D_{col}(z)= \frac{\sqrt{\sum_{i=1}^{2}
[col^{th}_i(z)-col^{obs}_i]^2}}{\sqrt{2}}.
\end{displaymath}

\noindent
We define the following criterion on the minimum distance
$D_{col}^{min}$:
{\small
\begin{eqnarray}
D_{col}^{min}&<&0.25 ~{\rm mag}.
\end{eqnarray}
}
\noindent
The application of the criteria (5--6) to the ACS catalogues produces
the sample is listed in Table~\ref{add-cand}, which consists of $2$
candidates in the northern field and $6$ candidates in the southern
one. Among the candidates in Table~\ref{add-cand}, we find the two
$z>4$ QSOs already listed in Table~\ref{new-cand}.  In Fig.
\ref{vi-iz} we plot the colors of the candidates listed in
Table~\ref{add-cand}.
\begin{figure*}[htp]
\centering
\includegraphics[width=12cm]{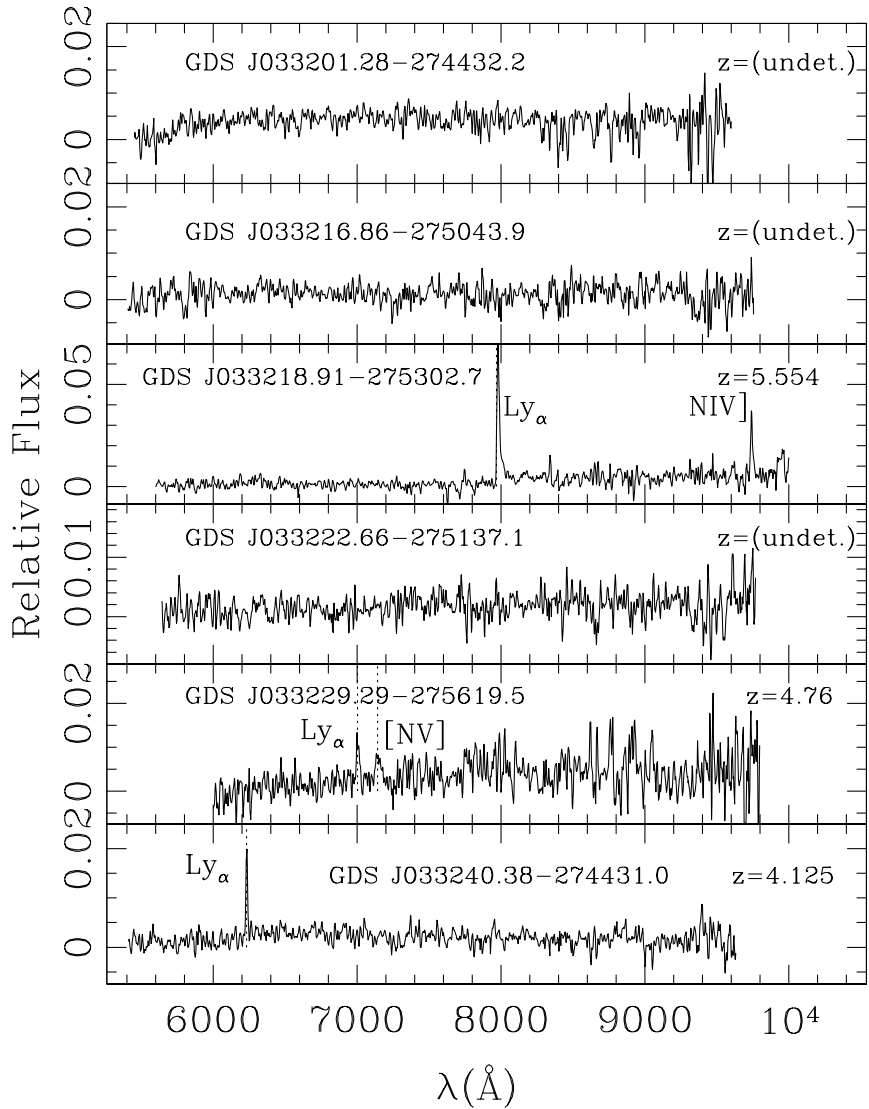}
\caption{Spectra of the morphologically+color selected QSO candidates listed in Table~\ref{add-cand}.
\label{spfig}}
\end{figure*}

A substantial fraction of these objects were observed during GOODS
spectroscopic surveys (Vanzella et al. 2006 for the CDF-S, Stern et
al. 2006 for HDF-N). The spectra of the CDF-S objects are shown in
Fig.~\ref{spfig}.  Two objects in the CDF-S ({\sc GDS
  J033240.38-274431.0} and {\sc GDS J033218.91-275302.7}) turn out to
be high-redshift galaxies, but they lack any evidence of AGN activity.
One of them ({\sc GDS J033218.91-275302.7}) shows an extremely strong
$Ly_{\alpha}$ feature in emission and the [NIV] intercombination line,
which, together with the absence of [NV], is suggestive of an H II
galaxy (Fosbury et al. 2003) at a redshift $z = 5.554$. The spectra of
{\sc GDS J033201.28-274432.2} and {\sc GDS J033216.86-275043.9} show
no emission lines. The quality of the spectra is not sufficient to
determine a redshift, but enough to exclude their being QSOs in the
redshift range $3.5 < z < 5.2$. The proposed object in HDF-N has been
observed in the framework of the GOODS project (D.Stern, private
communication). The resulting spectrum is inconclusive, but no sign of
AGN activity has been revealed.

On the basis of the morphological selection, no new QSO has been found
with respect to the X-ray criterion and no evidence gathered for an
incompleteness in the selection of Sect.~\ref{xsec}. The sample listed
in Table \ref{new-cand} is therefore considered a reliable estimate of
the total QSO space density at $z > 3.5$.
\begin{table*}
\begin{minipage}[t]{\columnwidth}
\caption{Morphologically+color selected QSO candidates}
\label{add-cand}
\centering
\renewcommand{\footnoterule}{}  
\begin{tabular}{cccccccccccc}
\hline
\hline
\multicolumn{10}{c}{High-redshift QSO candidates in the CDF-S.}\\
\multicolumn{2}{c}{Optical} & $z_{850}$ & $V-i$ & $i-z$ & $B-V$ &
 Class & Flux & & \multicolumn{2}{c}{spectroscopic}\\
RA & DEC & AB & AB & AB & AB & star & Radius &
redshift\footnote{Data in italics refer to the photometric estimate of the
  redshift under the assumption of QSO spectrum.} &
\multicolumn{2}{c}{identification} \\ 
{$3^h+ ~m~ s$} & {$-27^{\circ}+ ~'~``$} & (mag) & (mag) & (mag) &
(mag) & {(pixels)} & (pixels) & \\
\hline
 32 01.28& 44 32.2& 25.206& $2.19$& $0.15$& $>3$  &  $0.75$& $1.97$& {\it 4.85}& & \\ 
 32 16.86& 50 43.9& 25.190& $0.56$& $0.25$& $2.82$&  $0.92$& $1.39$& {\it 3.81}& & \\
 32 18.91& 53 02.7& 24.561& $2.56$& $0.63$& $>3$  &  $0.85$& $1.53$& 5.554 & EL--Galaxy & \citet{eros05} \\
 32 22.66& 51 37.1& 25.160& $1.49$& $0.47$& $>3$  &  $0.98$& $1.21$& --- & Star & \citet{eros06} \\ 
 32 29.29& 56 19.4& 24.984& $1.72$& $0.14$& $>3$  &  $0.99$& $1.31$& 4.759 & QSO & \citet{eros05} \\
 32 40.38& 44 31.0& 25.223& $0.53$& $0.01$& $3.10$&  $0.78$& $1.65$& 4.125& Galaxy & \citet{eros06} \\
\hline
\hline
\multicolumn{10}{c}{High-redshift QSO candidates in the HDF-N.}\\
{$12^h+ ~m~ s$} & {$62^{\circ}+ ~'~``$} \\
\hline
 36 47.96& 09 41.6& 23.761& $2.07$& $0.15$ & $4.81$ & $0.99$& $1.27$& 5.186 & QSO & \citet{Barger02}\\ 
 37 11.81& 11 33.5& 25.141& $1.51$& $0.11$ & $>3$   & $0.85$& $1.60$& {\it 4.700}& &\\ 
\hline
\end{tabular}
\end{minipage}
\end{table*}

\section{Estimating the luminosity function of high-z QSOs}
\begin{figure}[tp]
\centering
\includegraphics[width=8.5cm]{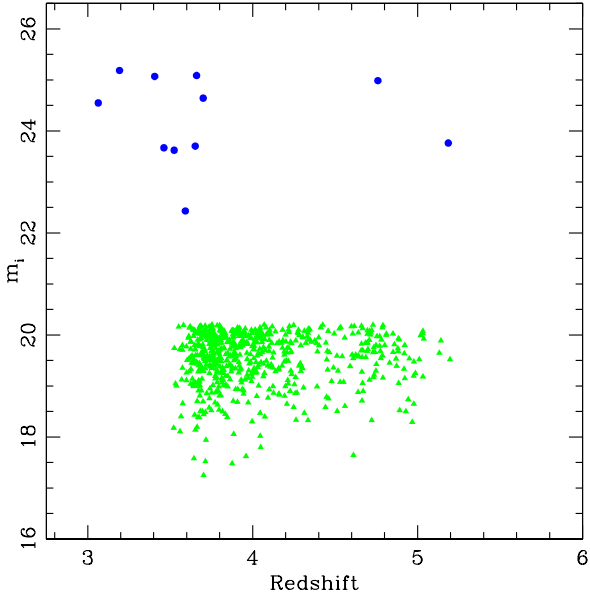}
\caption{Distribution of the observed QSOs (SDSS: triangles,
GOODS: circles) in the redshift - $m_i$ space. 
\label{ired}}
\end{figure}

In this section we describe the procedure we adopted in order to build
up the high-z QSO LF from the joint analysis of GOODS and SDSS
observations. The two samples are complementary in terms of the
surveyed area and the brightness limit and are characterized by
different photometric systems, selection criteria, and spectroscopic
completeness. In Fig.~\ref{ired} we show the position of the observed
SDSS and GOODS QSOs in the $z$-$m_i$ space (where $m_i$ refers to the
$i$ magnitude either in the GOODS or SDSS). It is evident from the
figure that the two surveys cover two non-overlapping regions of this
space.

\subsection{Simulated catalogs}
To combine the GOODS and SDSS information in order to estimate the QSO
LF, we adopted the approach of La Franca \& Cristiani (1997, see the
appendix for more details on the procedure).

We assume an LF of the form of a double power law:
\begin{displaymath}
\Phi(L_{145},z) =
\frac{\Phi^{\star}(L^{\star})}{(L_{145}/L^{\star})^{-\alpha}+(L_{145}/L^{\star})^{-\beta}}
\end{displaymath}
or, expressed in magnitudes:
\begin{displaymath}
\Phi(M_{145},z) =
\frac{\Phi^{\star}(M^{\star})}{10^{0.4(\alpha+1)(M_{145}-
M^{\star})}+10^{0.4(\beta+1)(M_{145}-M^{\star})}}.
\end{displaymath}
\noindent
We model a pure luminosity evolution (PLE)
\begin{displaymath}
L^{\star} = L^{\star}_{(z=2.1)} [(1+z)/3.1]^{k_z}
\end{displaymath}
or in magnitude:
\begin{displaymath}
M^{\star} = M^{\star}_{(z=2.1)} -2.5 \, {k_z} \, \log[(1+z)/3.1]
\end{displaymath}
and, alternatively, a pure density evolution (PDE) with a power law form
\begin{displaymath}
\Phi^{\star} = \Phi^{\star}_{(z=2)} \, [(1+z)/3]^{k_z}
\end{displaymath}
or, alternatively, with an exponential form
\begin{displaymath}
  \Phi^{\star} = \Phi^{\star}_{(z=2)} \, e^{{k_z}[(1+z)-3]}.
\end{displaymath}

Given a value for the break luminosity ($L^{\star}$ or $M^{\star}$),
the slopes of the double power law $\alpha$ and $\beta$, the
normalization $\Phi^{\star}$, and the redshift evolution parameter
$k_z$, we are able to calculate the expected number of objects from
the LF up to a given magnitude in a given area of the sky.  For each
object we extract a value of absolute $M_{145}$ and a redshift,
according to the starting parameters of the LF.  We then associate a
template spectrum to it, randomly chosen from our library (see
Appendix A). Using the template k-correction and colors at the
corresponding redshift, we simulate the apparent magnitudes in the
SDSS and ACS photometric systems. We also add the effect of
photometric errors in each band, estimated according to the analysis
of the photometric uncertainties as a function of the magnitude in the
DR3QSO and GOODS catalogues.  In this way we end up with mock
SDSS+GOODS catalogues and we then apply to them the selection criteria
paper I and R02, obtaining selected samples of simulated QSOs.

\subsection{Computing the QSO LF}
\label{compuLF}
The advantage of such an approach is related to being able to generate
and analyze a set of simulated SDSS+GOODS samples with the same
properties as the real ones in term of selection criteria,
completeness, color, and redshift distribution. We can associate
simulated catalogues to each starting set of LF parameters.  Using
multiple realizations for each parameter set, we are also able to
limit the statistical error in those simulations, which we can use to
estimate the goodness of the agreement of a given LF to the observed
sample, by directly comparing observed and simulated objects.

Following the approach described in detail by La Franca \& Cristiani
(1997), we divided the $z-m_i$ space into bins according to the
distribution of the observed objects and we defined a $\chi^2$
statistic by comparing the simulated and observed objects in each bin.
By using a minimization technique we can then compute the best-fit
parameters of the LF.

The functional form of our LF has five free parameters. In previous
studies they were estimated in various ranges of redshift.  Croom et
al. (2004) (hereafter Cr4) have found that the LF is approximated well
by a PLE in the redshift range between $0.4 < z < 2.1$.  We use their
estimated parameters at $z = 2.1$ as a starting point for our PDE and
PLE models.  We try various possibilities, differing in the number of
parameters fitted/fixed and in the type of evolution.  As a first
attempt we optimized only the evolutionary parameter $k_z$ (models
listed in Table~\ref{bfp} as 1, 2, 8, 9, 15, 16), then we fit two
parameters $k_z$ and $M_\star$ (models 3, 4, 5, 10, 11, 12, 17, 18,
19), and finally we optimized the three parameters $k_z$, $M_\star$
and the faint-end slope $\alpha$.

In a recent paper Richards et al. (2005, hereafter Ri5) compiled a
sample of quasars using imaging data from the SDSS and spectra taken
by the 2dF facility at the Anglo-Australian Telescope. Their data are
in good agreement with the 2QZ results of Cr4 at the bright-end, but
they require a steeper faint-end slope. We also considered models with
the faint-end slope taken from Ri5, $\alpha = 1.45$, and the other
parameters from Cr4 (models 2, 4, 9, 11, 16, 18).

Fan et al. (2001) found a shallower slope at high redshift for the
bright-end of LF with respect to Cr4 using SDSS commissioning data.
Richards et al. (2006) confirm these findings. At $z<3$ they estimate,
in agreement with Cr4, a bright-end slope $\beta \simeq -3.31$,which
progressively flattens to become $\beta \simeq -2.5$ at $z>4$. To take
this possibility into account, we also tested models fixing
$\beta=-2.58$ and fitting the other parameters (models 7, 14, 21).

\begin{figure*}[tp]
\centering
\includegraphics[width=12cm]{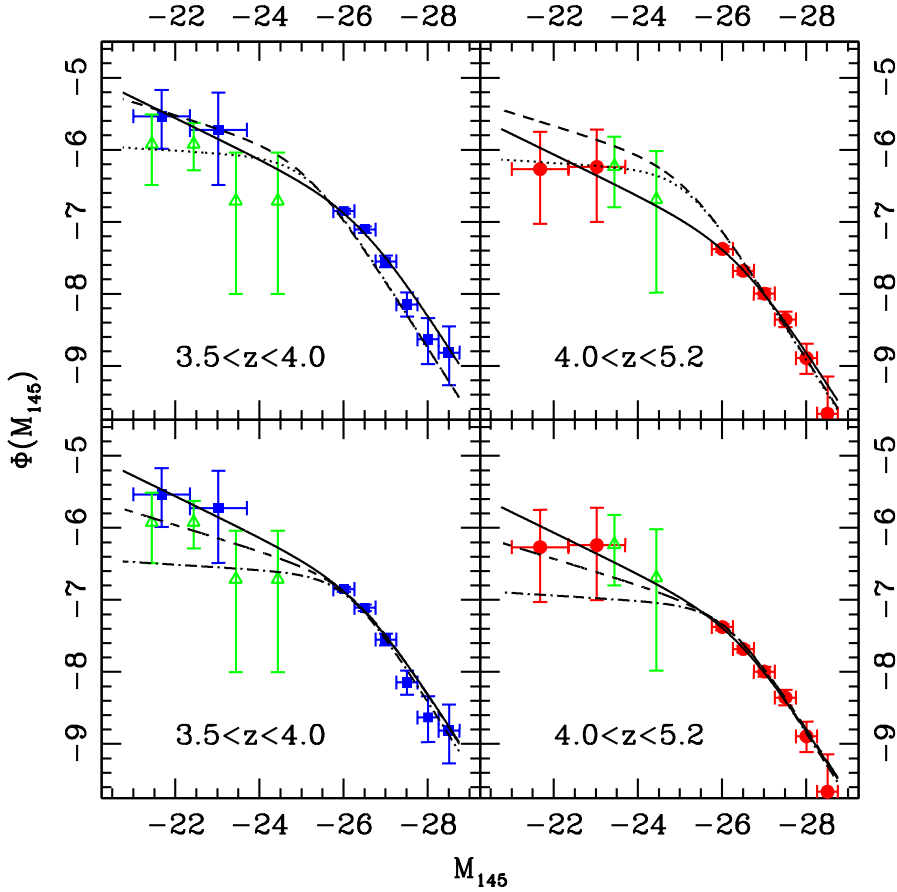}
\caption{Analytical fits to the high-z QSO LF: pure density evolution
  models with exponential evoution. In all panels the solid line
  shows, as a reference, the model with three free parameters (Nr.~13a
  in Table~\ref{bfp}). Upper Panel: models with one free parameter;
  dotted and dashed lines respectively refer to models with the Croom
  et al. (2004) and Richards et al. (2005) faint-end (Nr.~10 and~11).
  Lower Panel: models with two free parameters; dot-dashed and
  long-short dashed lines respectively refer to models with the Croom
  et al. (2004) and Richards et al. (2005) faint-end (Nr.~8 and~9).
  The position of the filled symbols was obtained by
  multiplying the values of model Nr.~13a by the ratio between the
  number of observed sources and the number of simulated sources. La
  Franca \& Cristiani (1997) demonstrated that this technique is less
  prone to evolutionary biases with respect to the conventional
  $1/V_{max}$ technique. Empty triangles refer to the results of
  the COMBO17 survey (Wolf et al. 2003).
  \label{plepke}}
\end{figure*}
\begin{figure*}[tp]
\centering
\includegraphics[width=12cm]{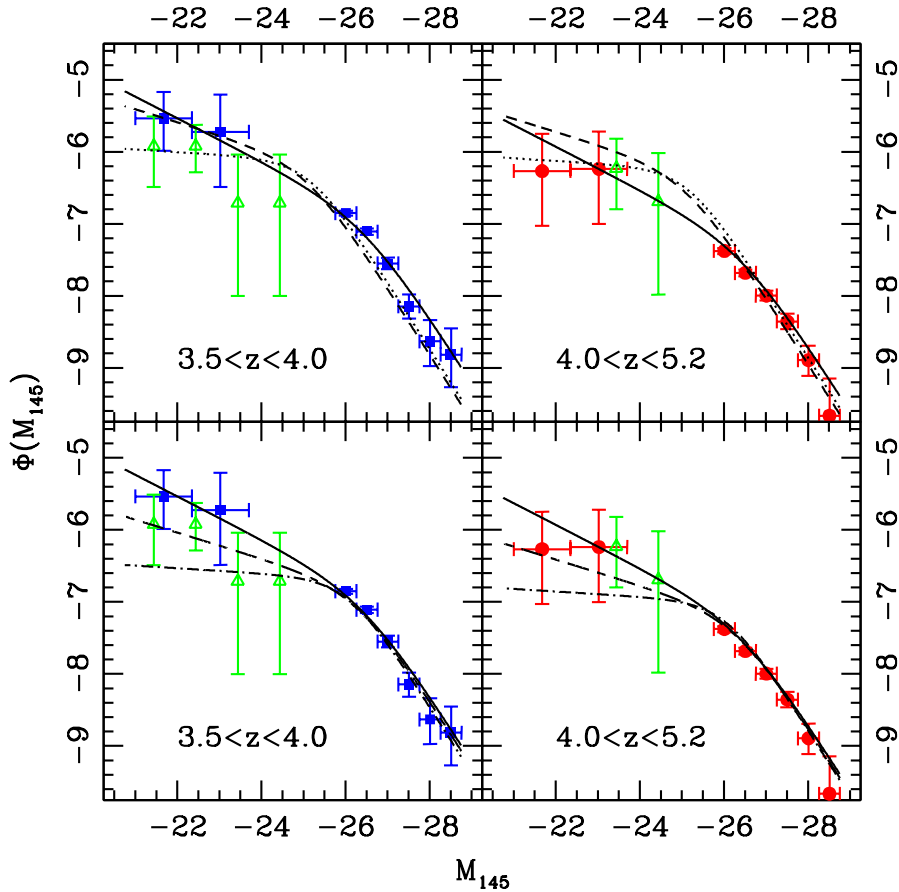}
\caption{Analytical fits to the high-z QSO LF: pure density evolution models 
with power-law evolution. Symbols have the same meaning as in Fig.~\ref{plepke}
\label{plepde}}
\end{figure*}

As a final statistical check of the results, we compared the simulated
and observed QSO distributions with a Kolmogorov--Smirnov
bidimensional test (see e.g. Fasano \& Franceschini 1987), producing
the probability $P_{KS}$\footnote{It is worth noting that in general
  the goodness of fit $\chi^2$-wise and $P_{KS}$-wise is correlated,
  but the best-fitting parameters are slightly different - though
  compatible within the uncertainties - when minimizing the $\chi^2$
  or maximizing the $P_{KS}$} listed in the last column of
Table~\ref{bfp}.

\subsection{Results and discussion}
Table~\ref{bfp} summarizes the results for all the models described in
the previous section, indicating for each of them the fixed
parameters, the fitted parameters, the corresponding value of the
reduced $\chi^2$, and the 2-D Kolmogorov--Smirnov probability
$P_{KS}$.  The best-fitting models are shown in Fig.~\ref{plepde}.
 
In general it turns out to be impossible to satisfactorily reproduce
the observed distribution of objects in the $m-z$ plane with PLE
models. On the other hand, PDE models show qualitatively different
results according to the number of fixed parameters and the assumed
evolution.  Models with one free parameter (1, 2, 8, 9, 15, 16) are
not able to reproduce the GOODS and SDSS surveys with the assumed
$M_\star = -25.116$.  The agreement becomes satisfactory for PDE
models where we fit both $k_z$ and $M_\star$ (4, 5, 10, 11, 12).  We
interpret this result as an indication of $M_\star$ evolution from $z
\sim 2.1$ to redshifts $3.5-4$. We also notice that models with a
steeper faint-end slope agree with observations better. This finding
is confirmed by models with three free parameters: in particular Nr.
13a, which shows a relatively high probability for a faint-end slope
as steep as $-1.71$.  Models with an exponential evolution of the
density tend to agree with the data better than do models with a
power-law evolution.

Models with a shallower slope of the bright-end of the LF (as
suggested by Fan et al. 2003) are not able to reproduce the observed
distributions: a bright-end slope as steep as Cr4 is required. This
result is at variance with Richards et al. (2006), we ascribe it to
the different estimates of the expected QSO colors between them and
the present work.  As explained in the Appendix, we have defined a QSO
template library starting from SDSS spectra in the redshift interval
$2.2 < z < 2.25$. The statistical properties of this library differ
from the template spectra used by the SDSS team to calibrate their
selection criteria. In particular the mean power-law slope of
continuum in our library is redder than the one assumed in Fan (1999).
As a consequence (Fig.~\ref{selcrit}), we estimate a lower
completeness of the SDSS color criteria with respect to Richards et
al. (2002), which is, in turn, the main reason for the discrepancy in
the estimate of the bright-end slope of the LF.

On the other hand, the present results also imply that a slope steeper
than Cr4 and similar to Boyle et al. (2000) is required for
reproducing the observed data for the faint end of the LF.

The present observations strengthen the conclusions of paper I about
the space density of high-z quasars and the joint evolution of
galaxies and QSOs.  Simple recipes, which are based on the assumption
that the QSOs shining strictly follows the hierarchical merging
structure of DMHs, grossly fail to predict the space density of high-z
QSOs. Models trying to account for the complexity of the baryonic
matter behavior with respect to DM have to be invoked (see paper I and
references therein for more details about the feedback effects from
AGN and stellar formation).

Following the method proposed by Barger et al. (2003), we can use our
best estimate for the LF (Nr.13a) to compute the QSO contribution to
the UV background at those redshifts.  We calculate the number of
ionizing photons per baryon produced in the redshift interval by the
observed AGN LF, and we convert the rest frame $ 1450$ \AA\ light to
the number density of ionizing photons using the form of the
near-ultraviolet spectrum given in \citet{Madau95}. We find a ratio
between ionizing photons and baryons on the order of $\sim 0.07$
between $3.5<z<4.0$ and of $\sim 0.02$ at $4.0<z<5.2$. This result is
similar to what Barger et al. (2003) found for $5 < z < 6.5$ QSOs,
confirming that the QSO contribution to the UV background is
insufficient for ionizing the IGM at these redshifts.
\begin{table*}
\begin{minipage}[t]{\textwidth}
  \caption{Best fit parameters of the LF of high-redshift QSOs}
\label{bfp}
\renewcommand{\footnoterule}{}
\centering
\begin{tabular}{cccccccc}
\hline
Model\footnote{The normalization of the Luminosity Function is in all cases $\Phi^{\star} = 1.67 \times 10^{-6} ~Mpc^{-3}$.} & Free   & Faint & Bright & $M^{\star}$ & $k_z$ & $\chi^2$ & $P_{KS}$ \\
Nr. & param.\footnote{Parameters for which no uncertainty is quoted have been fixed in the $\chi^2$ minimization.} & End   & End    \\ 
\hline

\multicolumn{8}{c}{Pure Density Evolution - Power-Law Evolution} \\
\hline
 1& 1 & -1.09            & -3.31 &  $-25.116$         & $-1.71 \pm 0.20$ & $3.2$  & $<1\%$\\
 2& 1 & -1.45            & -3.31 &  $-25.116$         & $-1.76 \pm 0.23$ & $3.2$  & $<1\%$\\
 3& 2 & -1.09            & -3.31 &  $-26.08 \pm 0.11$ & $-4.52 \pm 0.53$ & $1.38$ & $<1\%$\\
 4& 2 & -1.45            & -3.31 &  $-26.32 \pm 0.11$ & $-5.23 \pm 0.55$ & $0.99$ & $<5\%$\\
 5& 2 & -1.58            & -3.31 &  $-26.46 \pm 0.22$ & $-5.39 \pm 0.66$ & $0.95$ & $ 5.1\%$\\
 6& 3 & $-1.75 \pm 0.24$ & -3.31 &  $-26.48 \pm 0.26$ & $-5.52 \pm 0.94$ & $0.84$ & $ 15\%$\\
 7& 3 & $-1.98 \pm 0.71$ & -2.58 &  $-27.06 \pm 2.26$ & $-6.90 \pm 1.36$ & $1.5$  & $<1\%$\\
\hline

\multicolumn{8}{c}{Pure Density Evolution - Exponential Evolution} \\
\hline
 8& 1 & -1.09            & -3.31 &  $-25.116$         & $-0.46 \pm 0.10$ & $2.8$  & $<1\%$\\
 9& 1 & -1.45            & -3.31 &  $-25.116$         & $-0.40 \pm 0.15$ & $2.8$  & $<1\%$\\
10& 2 & -1.09            & -3.31 &  $-26.11 \pm 0.22$ & $-1.16 \pm 0.18$ & $0.87$ & $ 13\%$\\
11& 2 & -1.45            & -3.31 &  $-26.31 \pm 0.29$ & $-1.27 \pm 0.25$ & $0.63$ & $ 15\%$\\
12& 2 & -1.58            & -3.31 &  $-26.41 \pm 0.39$ & $-1.33 \pm 0.27$ & $0.62$ & $ 16\%$\\
13a\footnote{Parameters estimated through $\chi^2$ minimization.}
  & 3 & $-1.71 \pm 0.41$ & -3.31 &  $-26.43 \pm 0.45$ & $-1.37 \pm 0.46$ & $0.58$ & $ 5.5\%$\\
13b\footnote{Parameters estimated through $(1-P_{KS})$ minimization} 
  & 3 & $-1.49 \pm 0.24$ & -3.31 &  $-26.27 \pm 0.23$ & $-1.26 \pm 0.18$ & $0.65$ & $ 32\%$\\
14& 3 & $-1.93 \pm 0.31$ & -2.58 &  $-26.46 \pm 0.38$ & $-1.42 \pm 0.25$ & $1.3$  & $<1\%$\\
\hline

\multicolumn{8}{c}{Pure Luminosity Evolution} \\
\hline
15& 1 & -1.09            & -3.31 &  $-25.116$         & $-0.69 \pm 0.10$ & $2.8$  & $<1\%$\\
16& 1 & -1.45            & -3.31 &  $-25.116$         & $-0.72 \pm 0.11$ & $2.9$  & $<1\%$\\
17& 2 & -1.09            & -3.31 &  $-25.78 \pm 0.10$ & $-1.59 \pm 0.21$ & $1.9$  & $<1\%$\\
18& 2 & -1.45            & -3.31 &  $-25.82 \pm 0.10$ & $-1.76 \pm 0.23$ & $2.0$  & $<1\%$\\
19& 2 & -1.58            & -3.31 &  $-25.91 \pm 0.10$ & $-1.89 \pm 0.20$ & $2.0$  & $<1\%$\\
20& 3 & $-1.31 \pm 0.11$ & -3.31 &  $-25.73 \pm 0.25$ & $-1.56 \pm 0.31$ & $1.9$  & $<1\%$\\
21& 3 & $-1.31 \pm 0.35$ & -2.58 &  $-26.25 \pm 0.31$ & $-3.47 \pm 0.60$ & $1.0$ & $<1\%$\\
\hline

\end{tabular}
\end{minipage}
\end{table*}
\section{Summary}
In this paper we used the SDSS and GOODS databases to build an LF in
the magnitude interval $-21.0 < M_{145} < -28.5$ and in the redshift
interval $3.5 < z < 5.2$. In the first part of the paper we completed
the selection of the QSOs candidates in the GOODS fields carried out
by paper I.  We repeated the original analysis but using an improved
version of the optical catalogues, and discuss the results of the
spectroscopic follow-up of targets. The main conclusions of paper I on
the relatively low surface density of faint high redshift QSOs are
confirmed and strengthened.  We also explored the possible presence of
``X-ray faint'' AGNs, not detected in the A03 and G02 catalogues. This
additional search was based on a morphological analysis carried out
with the ACS optical images, selecting point-like candidates via a
direct comparison with the expected QSO colors. We followed up the
seven most promising candidates spectroscopically finding no
additional QSO among them and no evidence of any significant
population of faint X-ray high-z AGN with magnitude $M_{145} < -21$ in
our fields.

We then combined the QSO sample in the GOODS fields with the DR3QSO to
study the evolution of the QSO LF.  In order to understand the
systematics in the different samples, we exploited a method for
simulating the combined survey. This method is based on a library of
QSO template spectra, built up starting from observed SDSS QSO spectra
in the redshift interval $2<z<2.25$. Using this library we were able
to predict with great reliability the color distributions of observed
high-z QSO in the SDSS and GOODS photometric systems.  Assuming a
parameterization of the LF and its redshift evolution, we were then
able to simulate the expected object distribution in redshift and
apparent magnitude. From the statistical comparison between real and
simulated samples, we were able to quantitatively evaluate the
agreement of the assumed LF with the data.

We tried different LF models and our results point out that PDE models
show better agreement with observations than do PLE models, even if a
luminosity evolution of the magnitude of the break, $M^{*}_{145}$,
from $z=2.1$ to $z=3.5$ is required to obtain a good match.  The most
interesting result in terms of physical models is related to the faint
end slope of the LF. Parameterizations assuming a relatively steep
faint-end slope similar to Ri5 score a higher probability with respect
to those with a faint-end slope as flat as in Cr4.  If we also tried
fitting also the faint-end slope, we obtain a best-fit value even
higher than the Ri5 estimate.  A similar conclusion has been reached
by Hao et al.  (2005). Finally, we do not find any evidence of a
flattening at high-z of the bright-end of the LF with respect to the
Cr4 value, at variance with the results of Fan et al. (2003).

We used our best estimate for the LF to compute the QSO contribution
to the UV background at redshift $3.5<z<5.2$, and concluded that the
QSO contribution is insufficient for ionizing the IGM at these
redshifts.

When compared to physically motivated models, the present results
indicate a suppression of the formation or feeding of low-mass SMBHs
inside DMHs at high redshift.  Several processes can lead to this
phenomenon: photoionization heating of the gas by UV background
(Haiman, Madau \& Loeb 1999), feedback from star formation (Granato et
al. 2004), or the interaction between the QSO and the host galaxy. In
a recent paper, Monaco \& Fontanot (2006) studied this last
possibility in detail, providing hints of a new mechanism for
triggering galactic winds. In a forthcoming paper (Fontanot et al.
2006), this feedback mechanism will be inserted into a self-consistent
model for galaxy and AGN formation (Monaco, Fontanot \& Taffoni 2006):
the present results for the high-z LF will be used to constrain the
model and estimate the relevance of the joint feedback from supernovae
and AGN in shaping the redshift evolution of AGNs.
\begin{acknowledgements}
  We are greatly indebted and grateful to M. Dickinson, D.Stern,
  P.Tozzi, and the whole GOODS Team for the help in carrying out the
  observational program, enlightening discussions, and comments.
  Thanks!
\end{acknowledgements}

 \appendix{ 

\section{}
\subsection{QSO template library}
To build up a QSO template library, we decided to use high-quality 
SDSS QSO spectra at lower redshift. We needed a redshift
interval for which the SDSS sample had the highest possible level of
completeness and the continuum of the QSOs was sampled in the largest
possible wavelength interval from the $Ly_{\alpha}$ emission upward.
The interval $2.2 < z < 2.25$ was chosen on this basis. Among the
objects in the DR3QSO with $2.2< z < 2.25$, we selected the higher
quality spectra.  Our final sample consisted of 215 objects. We then
estimated the rest-frame spectrum of each object, extending the
continuum the blueward of the $Ly_{\alpha}$ line with the continuum
fitting technique derived from Natali et al.  (1998).  We defined
several ``continuum windows'' along the spectrum and fit the observed
fluxes with a power law. We used the resulting parameters to estimate
the intrinsic QSO spectrum (before IGM absorption) blueward of the
$Ly_{\alpha}$ line.  We used each spectrum in our library to compute
the theoretical QSOs colors at increasing redshift in the SDSS and ACS
photometric systems. To simulate the QSO spectra at different
cosmic epochs, we redshifted our spectra up to the redshift of interest,
then used a model for the IGM to simulate the IGM absorption.
\begin{figure}[tp] 
\centering
\includegraphics[width=8.5cm]{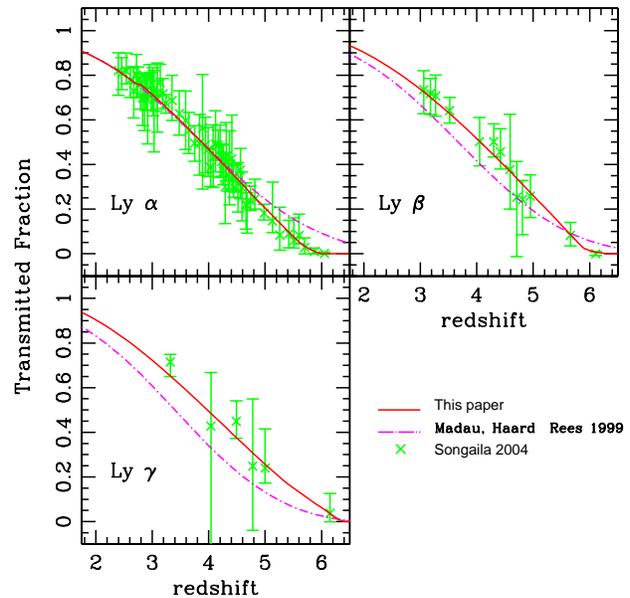} 
\caption{Comparison between different estimates of IGM absorption.
  Crosses (green in the electronic version) refer to Songaila (2004) observations; (magenta) dashed
  line refers to the prediction of the \citet{Madau95} model (applied to
  the QSO template spectrum of CV90), while the (red) solid line shows the
  applied IGM absorption.  
  \label{tau}}
\end{figure} 
To test the IGM model predictions at the redshift of interest,
we compared them with the observations of Songaila (2004). We applied the
IGM model to the CV90 template spectrum and computed the absorption
$\tau$ following the same procedure used by Songaila (2004) on real
spectra. The result is shown in Fig.~\ref{tau}.  Crosses refer
to the observed values of Songaila (2004), while the dashed line represents 
the prediction of the \citet{Madau95} model. It is evident from the figure that this model
overpredicts the $Ly_\alpha$ transmitted fraction at any redshift higher
than $4$, and underpredicts the $Ly_\beta$ and $Ly_\gamma$ at
redshift higher than $3$. We used the ratio between observations and
predictions to compute an empirical correction to the \citet{Madau95}
model at the corresponding wavelengths. The solid line in
Fig.~\ref{tau} represents the result when we applied our modified IGM
absorption model to the CV90 template spectrum.  

\subsection{QSOs color prediction}
We then computed the expected mean QSO colors using our template
library, providing a distribution of colors due to the different
continuum slopes and strengths of the emission lines in observed
spectra.

In Fig.~\ref{sdss_col} we show our results for the SDSS photometric
system compared to real colors in DR3QSO. The points refer to the
color of observed QSOs at different redshifts (see caption for more
details). In the left panel the dashed line refers to the R02 SDSS
selection criteria for $z > 3.5$ QSOs. Our predictions of QSO colors
are represented with solid lines (dotted lines are 5\% and 95\%
percentiles of color distribution in the template library. The
photometric errors are not accounted for.). The agreement is quite
good.

We can also predict the redshift evolution of QSO colors. In
Fig.~\ref{col_evo} we show the $g-r$, $r-i$, $i-z$ color (in the SDSS
photometric system) as a function of the redshift compared with
observed QSO colors.  The intrinsic scatter around the mean colors
(i.e. not taking photometric errors into account) is shown in
Fig.~\ref{sdss_col} (the dotted lines refer to 5\% and 95\%
percentiles of the color distribution). Similar results hold for the
GOODS survey.
\begin{figure*}[tp]
  \centering
  \includegraphics[width=8.5cm]{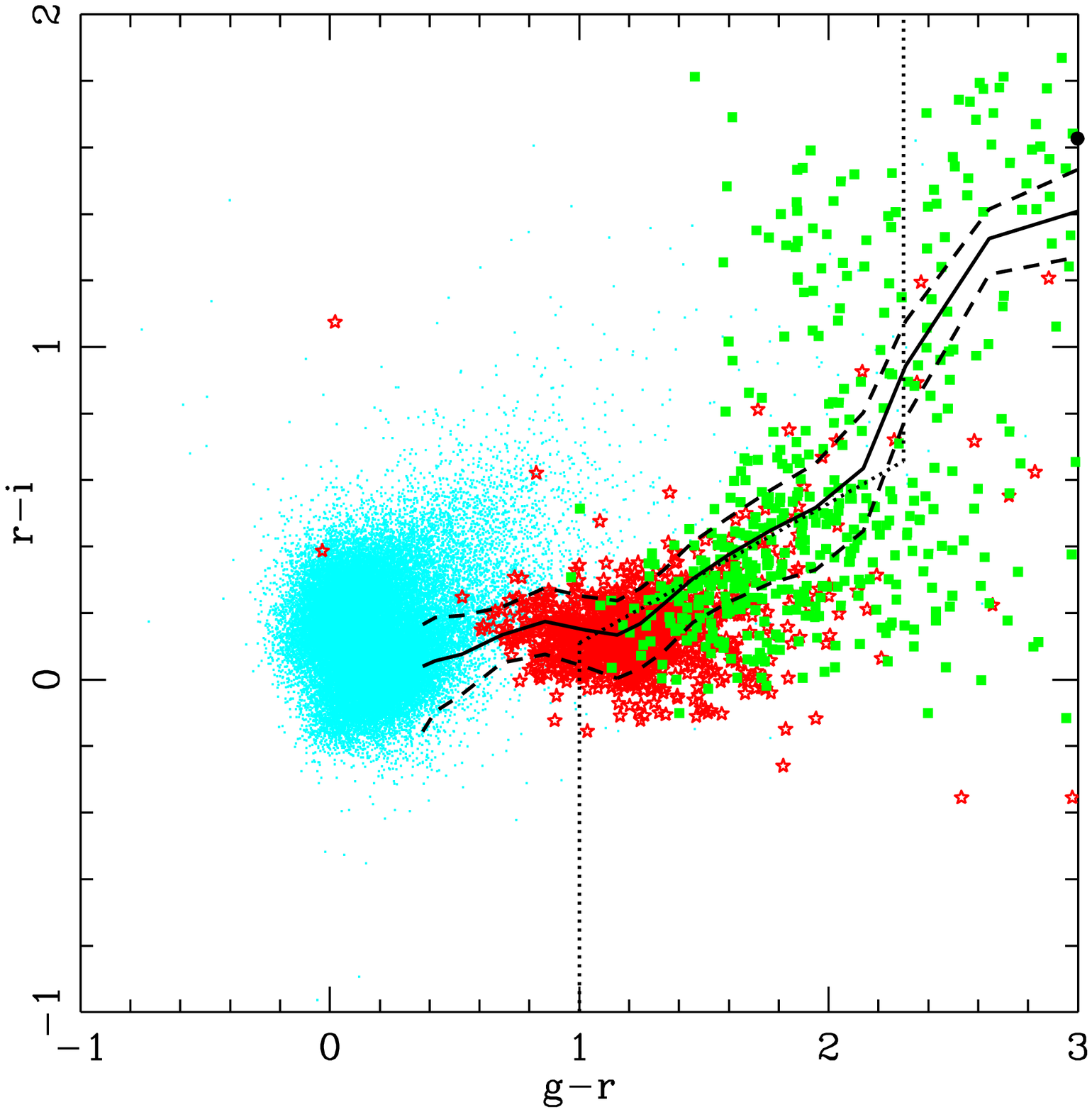}
  \includegraphics[width=8.5cm]{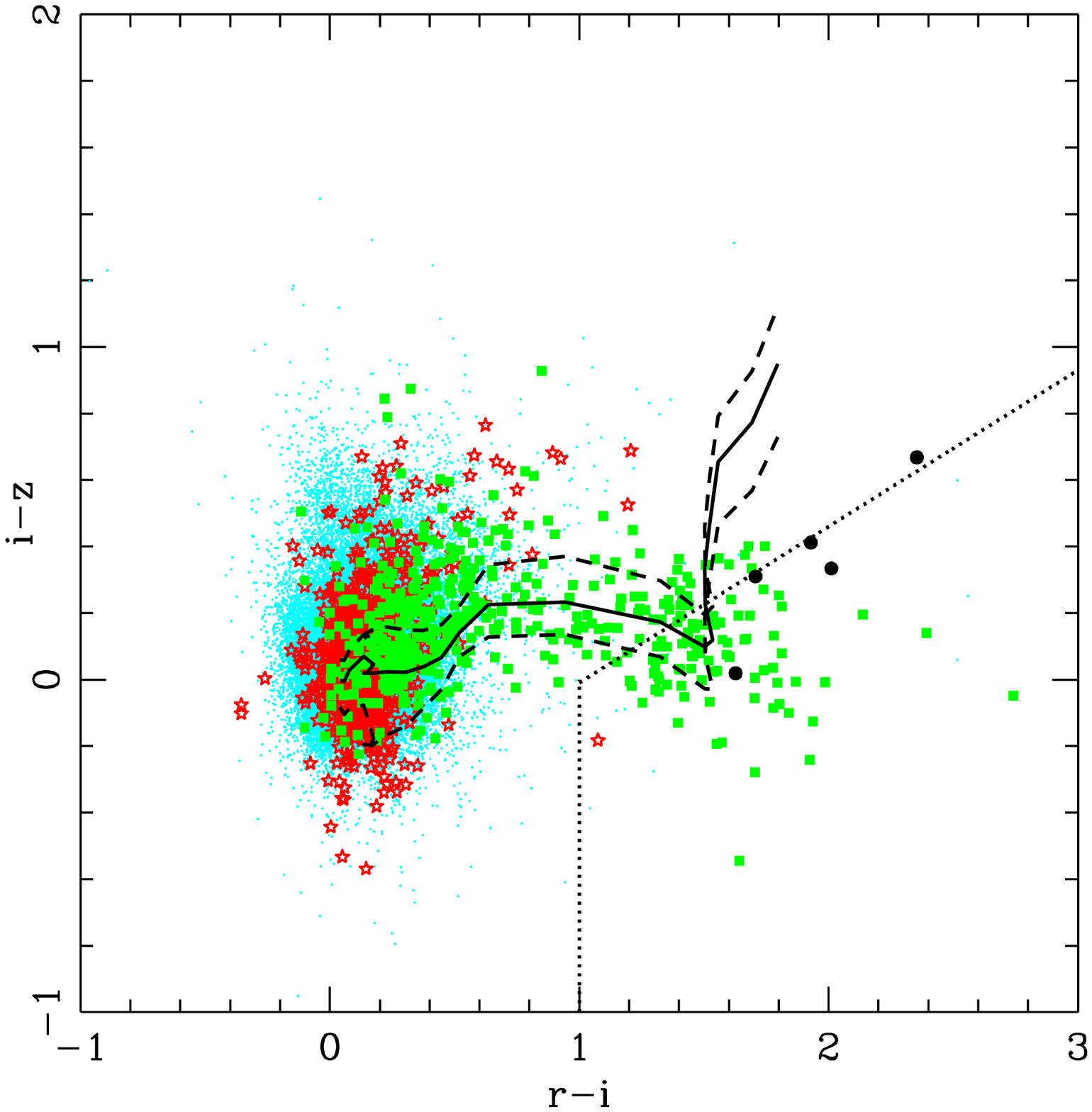}
\caption{Color diagrams for the confirmed QSOs in DR3QSO. Stars 
  (red in the electronic version) refer to objects with $3.5 < z < 4.0$;
  (green) squares refer to objects with $4.0 < z < 5.2$; filled circles 
  refer to objects with $z > 5.2$; (cyan) dots refer to objects with $z < 3.5$. 
  The solid line shows our prediction of QSO colors in the SDSS photometric
  system. Dotted lines are the 5\% and 95\% percentiles of color 
  distribution in the template library. The photometric errors are not 
  accounted for. Dashed line shows the selection criteria of Richards 
  et al. (2002) for $z>3.5$ QSOs.
  \label{sdss_col}}
\end{figure*}
\begin{figure*}[tp]
  \centering
  \includegraphics[width=8.5cm]{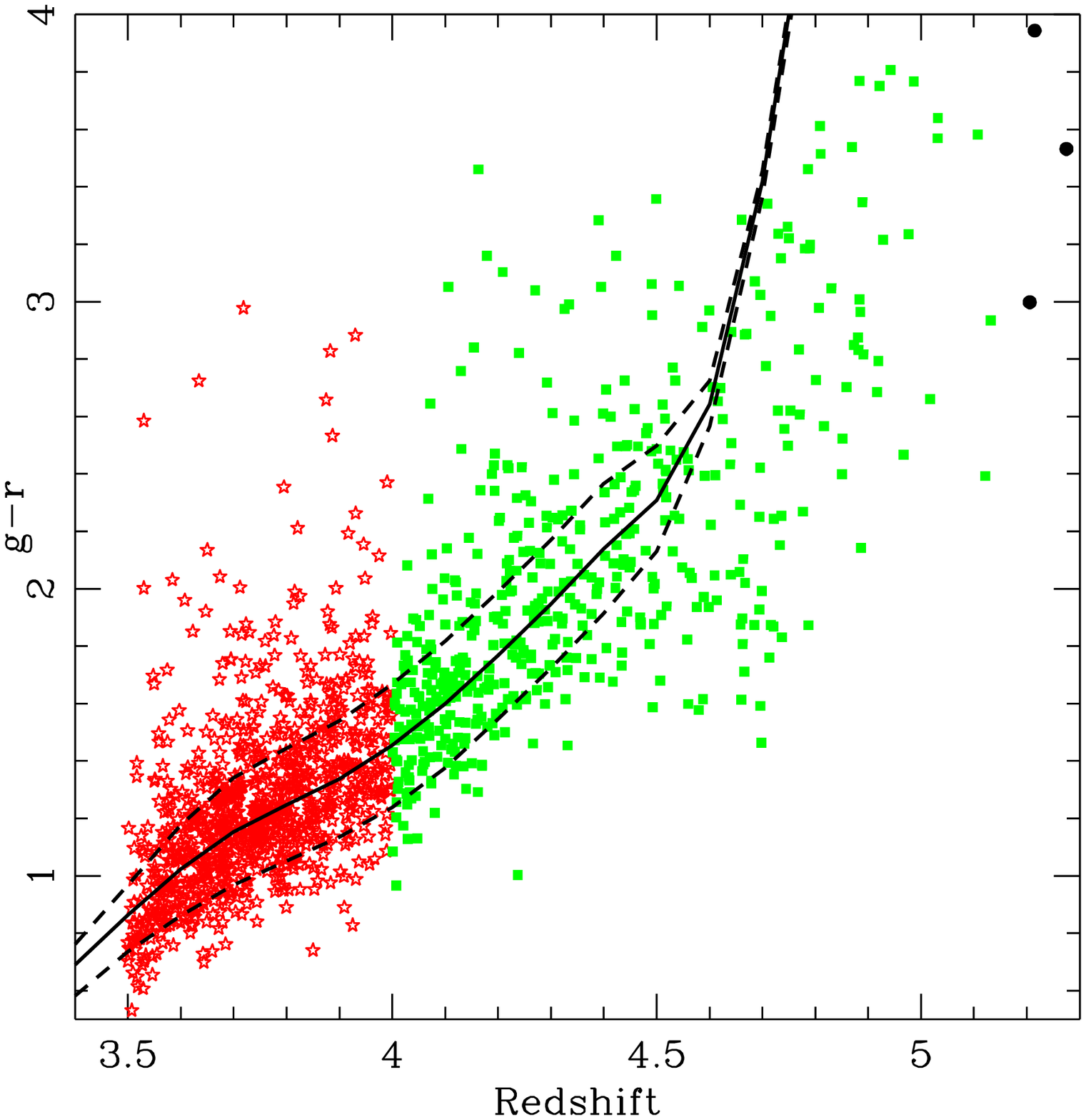}
  \includegraphics[width=8.5cm]{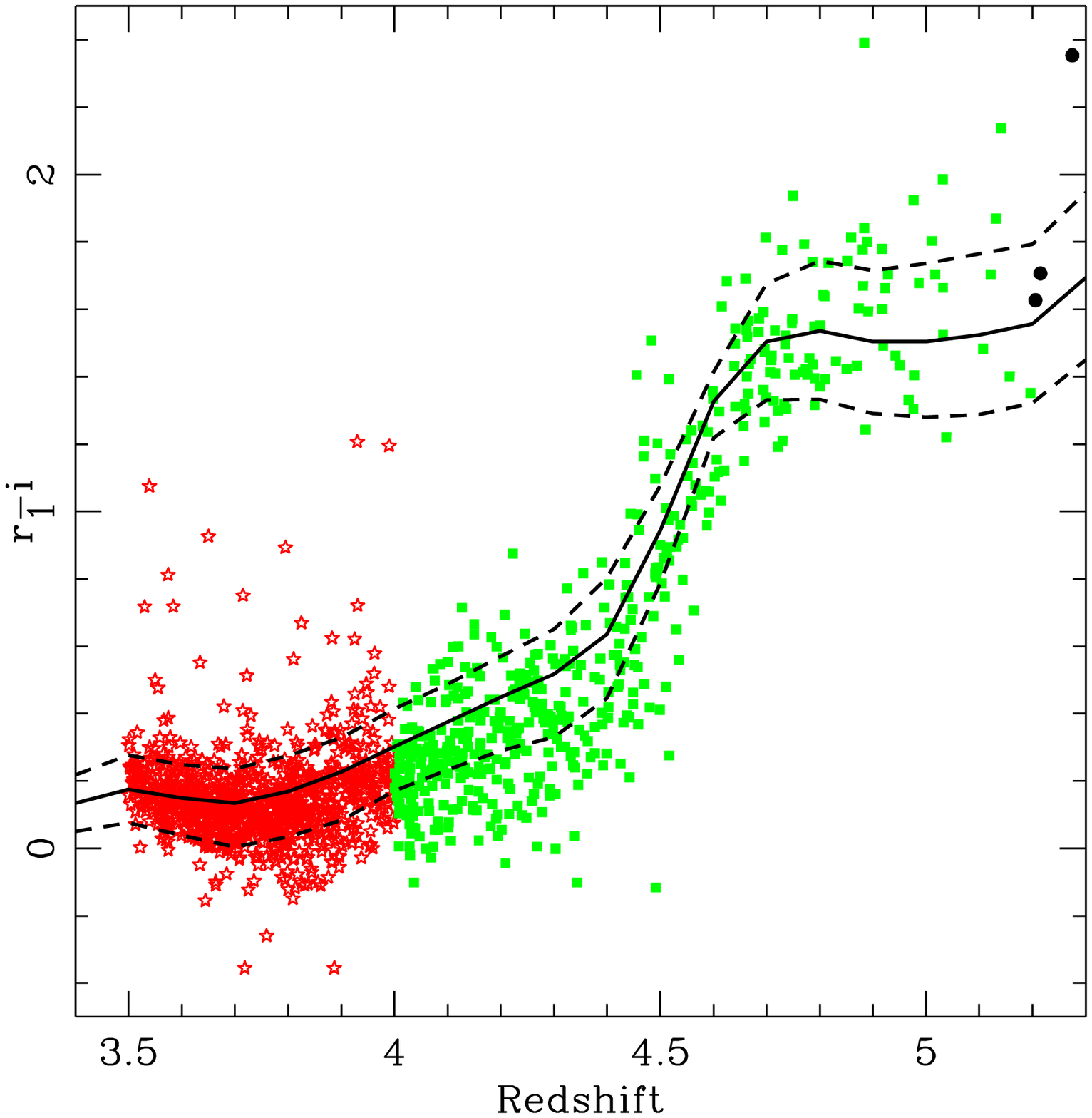}
  \includegraphics[width=8.5cm]{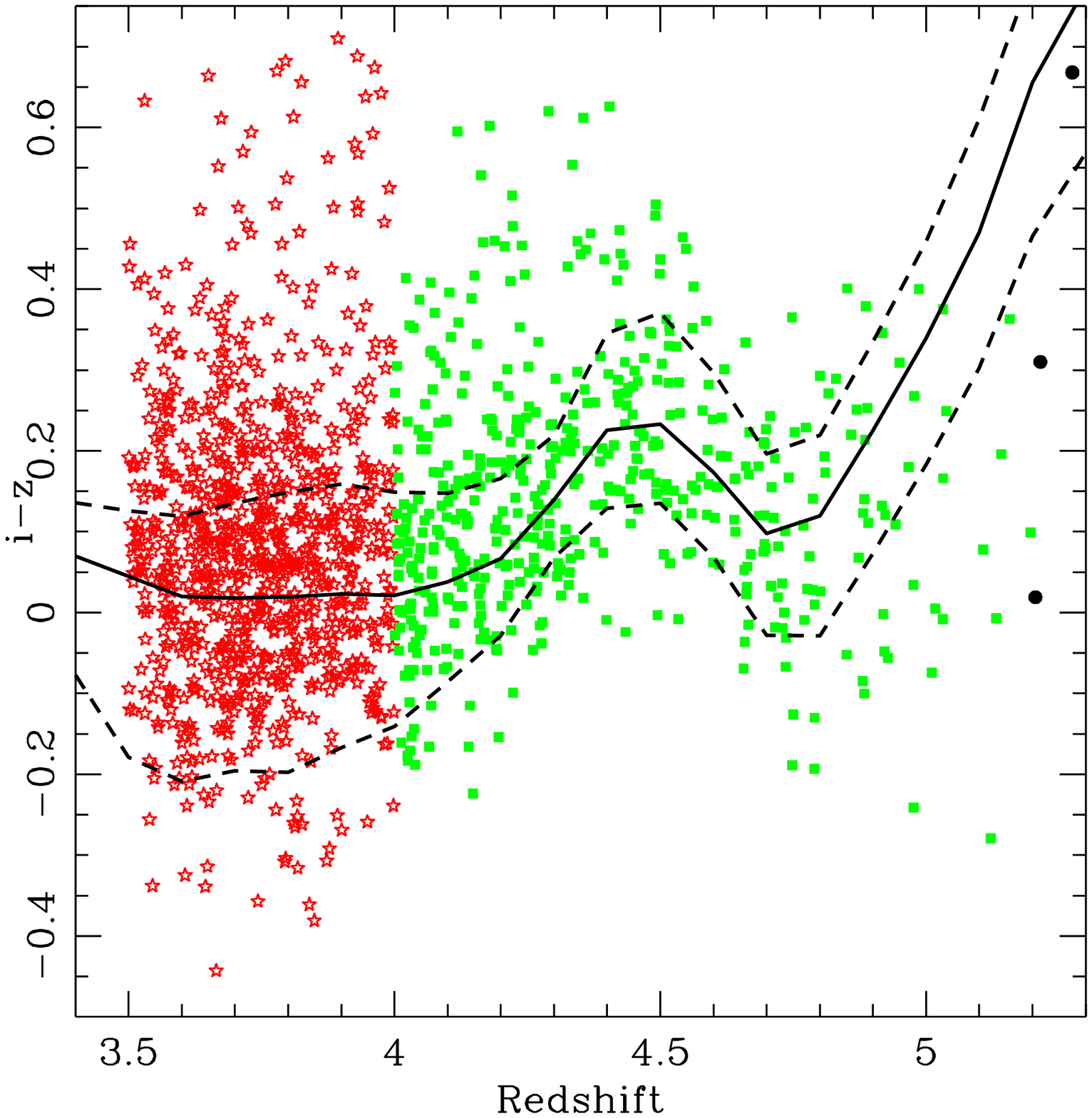}
\caption{Evolution of QSO colors with redshift in the SDSS photometric
  system. Symbols are the same as for Fig.~\ref{sdss_col}.
  \label{col_evo}}
\end{figure*}

\subsection{Completeness against selection criteria}
It is apparent from Figs.~\ref{sdss_col} and~\ref{col_evo} that we
are able to accurately predict the expected QSO colors as a function
of redshift.  Our template library can then be used to assess the
properties of the two independent samples. In particular we focused our
attention on the completeness of the samples selected using the Paper
I and R02 criteria.  In order to obtain a robust estimate of this
quantity we applied the selection criteria to our template library and
analyzed the fraction of QSOs recovered at various redshifts.  The
result is shown in Figure~\ref{selcrit} (left panel).  The SDSS
selection criteria shows a lower efficiency in selecting objects at $z
\sim 4.4-4.5$.  We tried to investigate this issue using the DR3QSO.
We considered the redshift distribution of all objects and compared it
with the redshift distribution of DR3QSO objects satisfying the R02
selection criteria.  The result is shown in Fig.~\ref{selcrit}
(right panel).  We found a dearth of $z \sim 4.0-4.5$ objects
satisfying the selection criteria with respect to the total sample.
To estimate its significance, we computed the number of object
in DR3QSO between $3.7<z<4.0$ and $4.5<z<5.2$ (677) and the number of
object in the same redshift interval satysfing R02 criteria (384). We
then rescaled the DR3QSO redshift distribution by the ratio between
these numbers, and we ended up predicting 210 QSOs with $4.0<z<4.5$.
Only 164 objects satisfying the R02 criteria were actually observed, a
discrepancy at $3-\sigma$ level. We also analyzed the same distribution
for the Fan et al. (2003) sample, finding a similar result.  The
result shown in Fig.~\ref{selcrit} (left panel) is different than
the analogous plot in Richards et al.  (2006), who estimated
the completeness of their criteria to be well above 90\% in the whole
range of redshift of interest.  This discrepancy is due to the
different QSO templates adopted.  In particular the SDSS selection
criteria are tailored on QSO template spectra whose mean continuum
slope $f_\nu \propto \nu^{-\gamma}$ is ``bluer'' ($\gamma = 0.5 \pm
0.3$ Fan, 1999) than the mean slope in our template library ($\gamma =
0.7 \pm 0.3$). The inferred completeness has a direct consequence on
the shape and evolution of the estimated QSO LF. Assuming the Richards
et al. (2006) completeness, the models with a shallow bright-end
(Nr.~7 and Nr.~14) score the highest probability, reproducing the
result by Richards et al. (2006).
\begin{figure*}[tp]
\centering
\includegraphics[width=8.5cm]{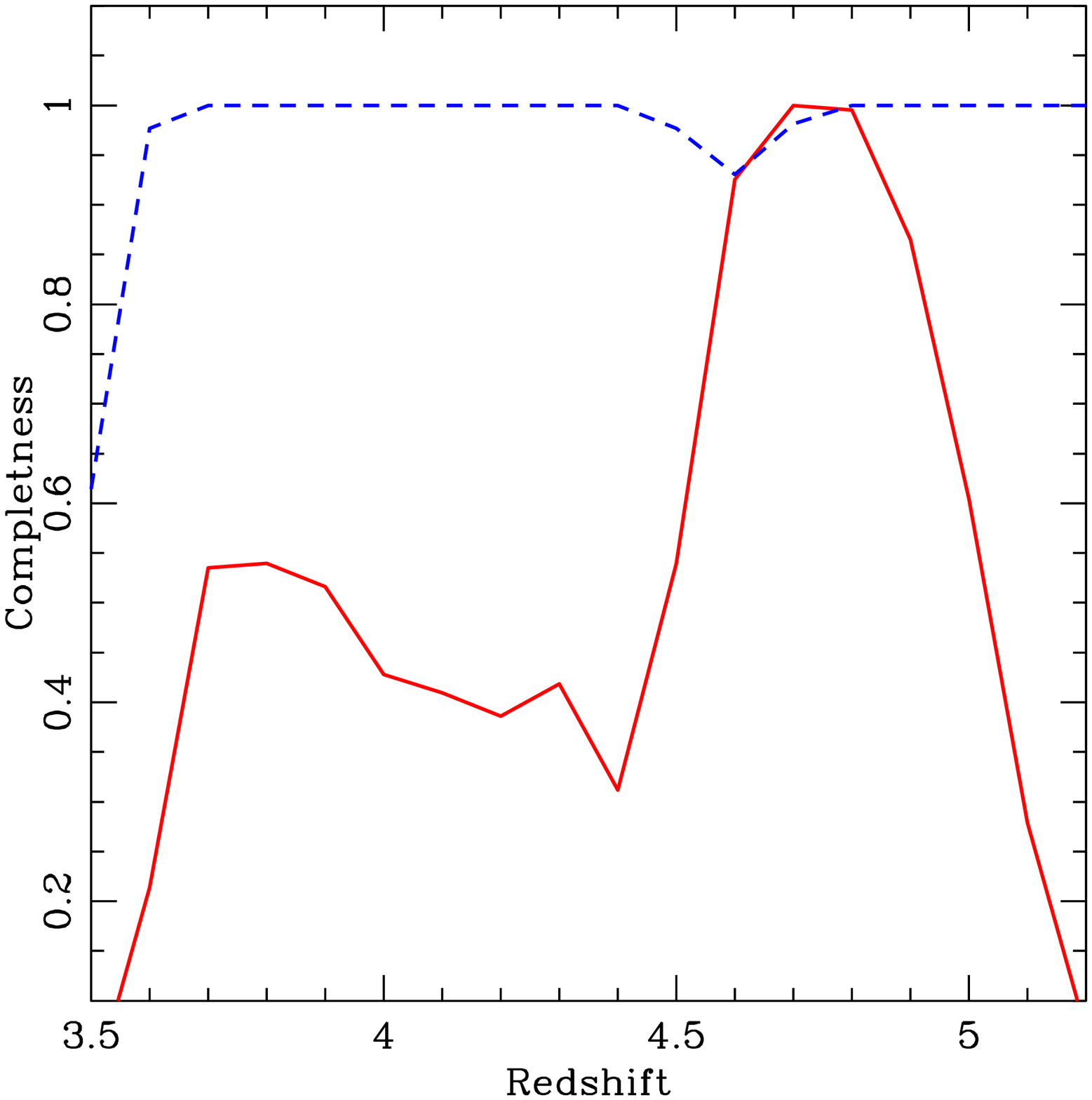}
\includegraphics[width=8.5cm]{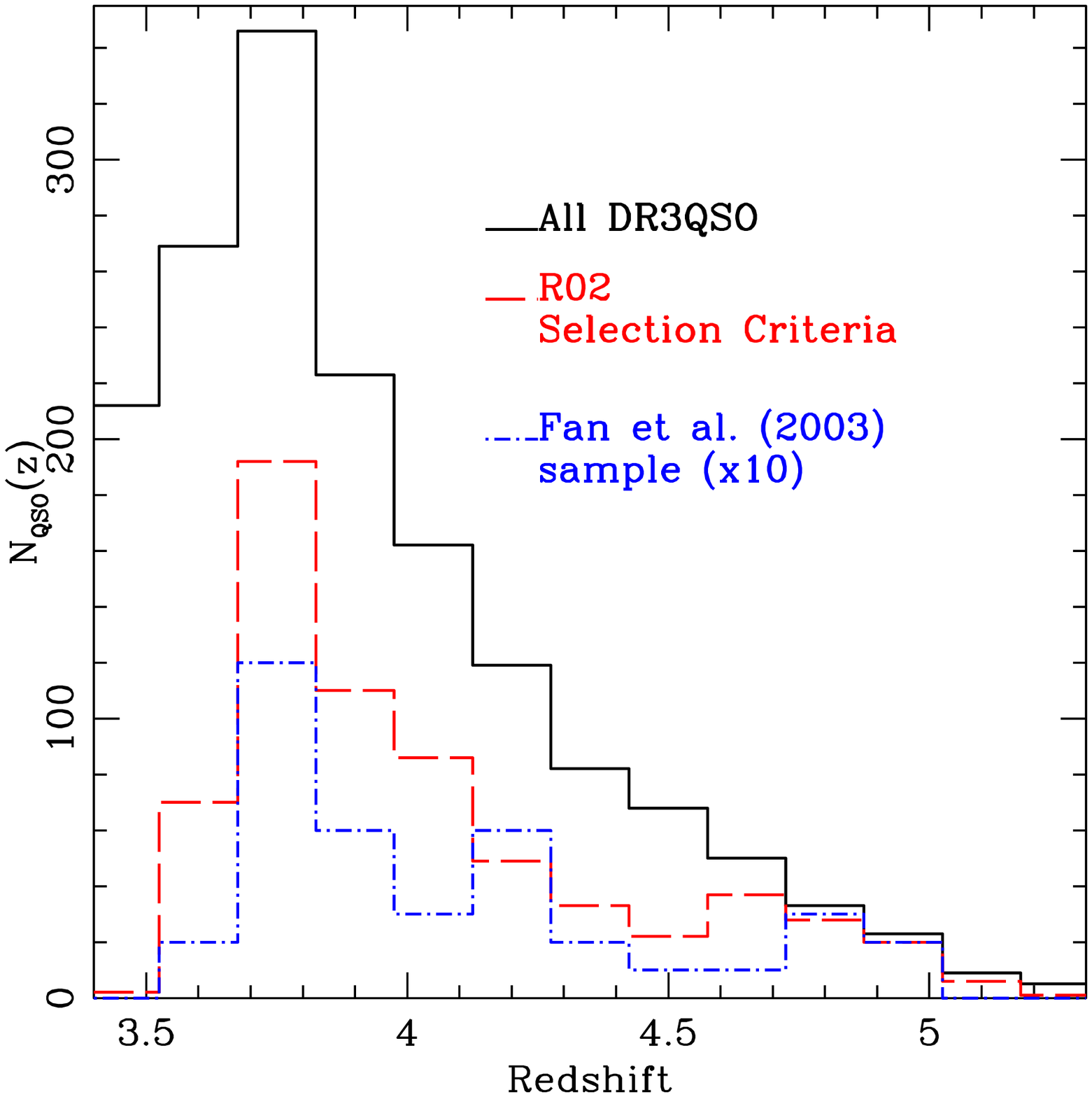}
\caption{Left panel: Completeness of selection criteria at various
  redshifts. The solid line refers to the R02 selection criteria
  for SDSS sources. The dashed line refers to paper I selection
  criteria in the GOODS fields. Right panel: Redshift distribution of
  DR3QSO sources (solid line). Dashed line represents the redshift
  distribution of DR3QSO sources satisfying the R02
  selection criteria. Dotted line represents the redshift distribution
  of the Fan et al. (2003) sample (multiplied by a factor 10 for graphical
  reasons) taken out of the SDSS commissioning data.
  \label{selcrit}}
\end{figure*}
}

\end{document}